\documentclass[article, aps,preprint,preprintnumbers,amsmath,amssymb,nofootinbib]{revtex4-1}
\usepackage{epsfig}% Include figure files
\usepackage{dcolumn}% Align table columns on decimal point
\usepackage{graphicx}
\graphicspath{ {images/} }
\usepackage{bm}
\usepackage{hyperref}
\usepackage{color}  % colored text to highlight revisions 
\usepackage{setspace}
\usepackage{appendix}
\usepackage{mathtools}

\newcommand{\vel}{\mathbf{u}}
\newcommand{\B}{\mathbf{B}}

\newcommand{\pd}{\partial }

\definecolor{Gray}{gray}{0.85}
\definecolor{LightCyan}{rgb}{0.88,1,1}
\begin{document}

\title{Scaling of reconnection parameters in magnetic island coalescence: Role of in-plane shear flow}

\author{Jagannath Mahapatra\textsuperscript{1,2}}
\email{jaga.physics94@gmail.com}
\email{jagannath.mahapatra@ipr.res.in}
%\author{Arkaprava Bokshi\textsuperscript{1}}
\author{Rajaraman Ganesh\textsuperscript{1,2}}
\email{ganesh@ipr.res.in} 
\author{Abhijit Sen\textsuperscript{1,2}}

\affiliation{\textsuperscript{1}Institute for Plasma Research, Gandhinagar, Gujarat-382 428, India}
\affiliation{\textsuperscript{2}Homi Bhabha National Institute, Training School Complex, Anushaktinagar, Mumbai 400094, India}
%
%\author{M. Jagannath}%\textsuperscript{1}}
%\email[ E-mail me at : ]{jaga.physics94@gmail.com/jagannath.mahapatra@ipr.res.in}
%\affiliation{Institute for Plasma Research, Bhat, Gandhinagar, 382428, India}
%\author{ R. Ganesh}%\textsuperscript{1}} 
%\affiliation{Institute for Plasma Research, Bhat, Gandhinagar, 382428, India}

\begin{abstract}
A 2D incompressible viscoresistive-MHD model [Mahapatra et al., Phys. Plasmas 28, 072103 (2021)] is used to study the scaling of reconnection parameters in the magnetic island coalescence problem under two interesting scenarios. Firstly, the effect of changing island half-width at a fixed system size is investigated. As the island half-width increases, the total magnetic flux content of the islands increases resulting in an increase in upstream magnetic field, upstream velocity field and unnormalized reconnection rate. However, the downstream magnetic field, current sheet length and normalized reconnection rate ({normalized to the upstream magnetic field and upstream Alfv\'enic velocity}) remains independent of it. Interestingly, the reconnection rate is found to be different from the upstream to downstream velocity ratio as well as from the aspect ratio of the current sheet, as opposed to the findings of the Sweet-Parker model. Secondly, the in-plane shear flow effects are studied, keeping the island width and system size fixed. Here thickness and length of the current sheet, the upstream magnetic and velocity field components, reconnection rate and time, current sheet inclination angle with shear flow length scale and amplitude are calculated. Interestingly, the inclination angle of the current sheet and the diffusion region are found to be different and the differences are more in stronger shear flows. These results are significantly different from the Harris sheet setup with shear flow. %The results have potential application in the Earth magnetospheric observations and fusion plasma experiments.
\end{abstract}
\maketitle

%Using a 2D viscoresistive incompressible MHD model, influence of two important parameter on coalescing magnetic islands and associated reconnection mechanism is studies. First parameter is the island half-width, intrinsic to Fadeev equilibrium,
%\newpage
\section{Introduction}
\indent
Magnetic reconnection is one of the major causes for many eruptive processes \cite{biskamp2000book, priest2000magnetic, Yamada2010review} observed in solar atmosphere \cite{priest2002review}, earth and other planetary magnetospheres \cite{HesseCassak2020review}, fusion plasma devices \cite{zweibelYamada2009review}, etc, that release a significant amount of magnetic energy of the system in the form of plasma bulk flow, thermal heating and nonthermal particle acceleration by reconfiguring the global magnetic field topology that leads to a lower energy state of the plasma system. The local magnetic null points are the primary reconnection sites where thin current sheets (CS) can form. The Alfv\'enic frozen-in condition breaks down inside these CS either due to collisional or collisionless processes, allowing magnetic field lines to break and(or) reconnect. {Particularly}, in collisional plasmas, the aspect ratio of the reconnecting CS satisfies $\delta/l >> d_i/L$ where $d_i$ is the ion skin depth given by $d_i = c/(\omega_{pi}L)$, $L$ is the characteristic length scale, $c$ is the light speed in vacuum, $\omega_{pi}$ is ion plasma frequency, $\delta$ is the thickness and $l$ is the CS length. Within this limit, one can assume a resistive-MHD model to be valid, {generating a Sweet-Parker (SP) \cite{parker1957sweet,sweet1958} like CS, and the reconnection rate along with the associated CS properties follow SP scaling} i.e. $\sim S^{-1/2}$ where the Lundquist number $S = L v_A/\eta$, $v_A$ is Alfv\'en velocity and $\eta$ is electrical resistivity of plasma. {Furthermore, the downstream flow $v_{out} \sim v_A$, the upstream flow $v_{in} \ll v_A$, {CS length $l$ is independent of $S$} and CS thickness $\delta \sim S^{-1/2}$}. However, for very large values of Lundquist number $S$ values, as in the case of Earth's magnetosphere and Solar atmospheric plasmas, the CS attains a kinetic scale thickness ($\delta \simeq l \sim d_i$) and the reconnection rate is found to be independent of resistivity, inline with many in-situ observations. \\%In the present work, we will confine our study to resistive-MHD only. \\
\indent In the past, most of the analytic and numerical work on magnetic reconnection studies have used a Harris sheet set-up \cite{harris1962} for both collisional and collisionless plasmas. {The initially supplied CS can be susceptible to various resistive instabilities {\cite{BirnHesse2007, cassak2011, Hosseinpour2018}} that lead to SP reconnection}. Magnetic island-coalescence (MIC) problem \cite{finnKaw1977, pritchett1979coalescence,biskamp1980, biskamp1982, Rickard1993, knoll2006POP, knoll2006PRL, BirnHesse2007, pritchett2007kinetic, karimabadi2011flux, stainer2015POP, stainer2015PRL, Jonathan2015, stainer2017, BardDorelli2018, mahapatra2021, murtas2021coalescence} that uses a {Fadeev equilibrium \cite{fadeev1965})} i.e. a 1D chain of current filament offers an alternate setup for reconnection study, has several advantages over Harris set-up. For example, in Harris equilibrium, the reconnecting CS is a part of an initially-supplied CS and is not coupled to the large scale magnetic flux source. Further, to initiate reconnection, the initial CS either needs to have a critical aspect ratio or requires an external driving force \cite{BirnHesse2007}. In MIC problem however, the reconnection is self-driven as the attractive Lorentz force between the parallel current carrying filaments drives the reconnection and forms a reconnecting CS at the $X$-point between them; {a scenario very much similar to many natural reconnecting systems \cite{zhao2016coalescence,zhao2019, feng2019observations, browning2014, browning2015, Song2012, wang2016coalescence}}. {Moreover, unlike the assumptions of SP-model in the Harris set-up, magnetic islands have a finite magnetic flux supply that controls the physical length of the reconnecting CS.} Hence, the MIC problem has been studied extensively to understand the physics of micro-scale CS and reconnection time scales using different numerical models such as resistive-MHD \cite{pritchett1979coalescence, biskamp1980, biskamp1982, Rickard1993, knoll2006POP, BirnHesse2007, mahapatra2021}, Hall-MHD \cite{DorelliBirn2003JGR, knoll2006PRL, Jonathan2015, stainer2015POP, stainer2017, BardDorelli2018}, EMHD \cite{dorelli2001emhd, chacon2007emhd}, two-fluid \cite{stainer2015POP}, hybrid model \cite{stainer2015POP, stainer2015PRL, Jonathan2015, stainer2017}, kinetic (Particle-In-Cell) model \cite{pritchett2007kinetic, karimabadi2011flux, stainer2015PRL}. Using some of these models, the influence of different plasma parameter such as guide field \cite{stainer2015POP, stainer2017}, compressibility \cite{BirnHesse2007}, system size \cite{karimabadi2011flux, stainer2015PRL, Jonathan2015, BardDorelli2018} on the evolution of MIC have been studied in detail. Recently, the role of in-plane shear-flow \cite{mahapatra2021,mahapatra2020FEC} on MIC process has been reported. In incompressible resistive MIC, two different scaling regimes with respect to $S$ has been suggested \cite{biskamp1980, biskamp1982, knoll2006POP}: a transient regime where the reconnection rate is insensitive to $S$ and a higher $S$ value regime where the SP scaling is followed. Similar scaling with compressibility effect is reported in Ref. \cite{Rickard1993} except that the downstream flow is found to be insensitive to $S$. Likewise, scaling with island wave-number (inverse of island wavelength $a_B$, see Sec. \ref{simulation-details} for definition), a proxy to island system size, has also been reported using different numerical models. In resistive-MHD, MIC follows SP scaling i.e. $\sim S_{a_B}^{-1/2}$ where $S_{a_B}^{-1/2} = a_B v_A/\eta$ \cite{Jonathan2015}. Further, in collisionless plasmas, the system size is compared with ion diffusion scale $d_i$, and the scaling with $a_B$ is found to be subtle: $\sim (a_B/d_i)^{-\alpha}$ ($\alpha$ ranging from 0.25 to 0.8 for Hall-MHD to PIC models \cite{stainer2015PRL}). In all the above studies, the $\epsilon$ parameter in Fadeev equilibrium \cite{fadeev1965} that controls island half-width or total flux content inside the island, is taken to be constant. {Following the analytical work by Finn and Kaw \cite{finnKaw1977},} Pritchet and Wo \cite{pritchett1979coalescence} have reported an increase in growth rate of coalescence instability as $\epsilon$ increases along with stabilizing effects of conducting walls, for a fixed $a_B$ value. Furthermore, Birn and Hesse \cite{BirnHesse2007} have found a SP like scaling of reconnection rate and driving $E_y$  with $S$ for four different $\epsilon$ values. However, other important parameters such as CS properties/aspect ratio, upstream/downstream magnetic field and flow components, magnitude of pressure pile-up for full range of $\epsilon$ ($0 \le \epsilon \le 1$) have not been studied. In the first part of this report we attempt to fill the above gap in literature. Moreover, this study will be helpful in understanding the competition between pressure pile-up and coalescence inducing Lorentz force in the MIC problem.\\
\indent Externally applied or internally generated shear flows are also found to influence the reconnection process in a Harris current sheet. In collisional magnetized plasma, super-Alfv\'enic antiparallel shear flows are susceptible to MHD-Kelvin-Helmholtz-instability (MHD-KHI) and suppress the current driven instabilities such as resistive tearing mode instability (TMI) \cite{cassak2011}, plasmoid instability \cite{Hosseinpour2018}, and double tearing mode instability \cite{Arghyadeep2021}. In collisionless plasmas, the MHD-KHI and TMI can coexist in sub-Alfv\'enic shear flows such that the reconnection can be driven by the interacting vortices (vortex-induced-reconnection). Shear flow effects on coalescing magnetic island evolution and associated reconnection rate are recently reported \cite{mahapatra2021}, suggesting a suppression of MHD-KHI in super-Alfv\'enic shear flows by the IC instability and is insensitive to shear flow scale length. The external shear flow is found to generate a secondary shear flow on both sides of CS that is responsible in affecting the reconnection rate. However, only qualitative effects on the reconnection rate have been reported. In continuation to the work reported in Ref. \cite{mahapatra2021}, here we report a quantitative study of shear flow effects on CS properties and other reconnection parameters. These details comprise of the second part of this report.  \\
\indent The paper is organized as follows. In Sec. \ref{simulation-details}, details of the simulation model and initial profile are discussed. Section \ref{Results} discusses the influence of two important parameters on the MIC problem, (1) effects of magnetic island half-width in Subsection \ref{size-study} and (2)  effects of in-plane antiparallel shear flows on island evolution and other reconnection parameters in Subsection \ref{shear-flow-study}. Finally, important findings of this study are summarized in Sec. \ref{summary}.

\section{Simulation details}\label{simulation-details} 
We use the BOUT++ \cite{dudson2009bout++,dudson2015bout++} framework to numerically solve the viscoresistive-RMHD equations \cite{mahapatra2021,knoll2006POP} in a 2D Cartesian geometry ((z-x) plane). {The assumption $\partial/\partial y = 0$ can be ensured in the presence a strong guide field along the out-of-plane direction, although it never appears in the model equation of vorticity - vector potential ($\omega$-$\Psi$) formalism. The governing equations are},
\begin{align}
	&\nabla \cdot \vel = 0,\\
	&\frac{\pd \omega_y}{\pd t} = \left[\varphi,\omega_y \right]  - \left[\Psi,J_y \right] + \hat{\nu} \nabla^2 \omega_y, \label{eqn:omegay-2d}\\
	&\frac{\pd \Psi}{\pd t} = \left[\varphi,\Psi \right] - \hat{\eta}(J_y -J_{y0}).  \label{eqn:psi-2d}
\end{align}
{The above equations are normalized to Alfv\'enic units: length $L$ to the system length $L_z$, velocity to Alfv\'enic velocity $v_A = B/\sqrt{\mu_0 \rho}$ (free space permeability $\mu_0 = 1$, normalized density $\rho= 1$ is assumed) and time $t$ to Alfv\'enic time $t_A = L/v_A$. In the equations,} vorticity $\omega_y = (\nabla \times \vel) \cdot \hat{y} = -\nabla^2\phi$, $\phi$ is the velocity stream function, in-plane velocity $\vel = \nabla \phi \times \hat{y}$, current density $J_y = (\nabla \times \B) \cdot \hat{y} = -\nabla^2\Psi$, $\Psi$ is the magnetic flux, magnetic field $\B = \nabla\Psi \times \hat{y}$, $\hat{\eta}$ is normalized resistivity, $\hat{\nu}$ is normalized viscosity, $\nabla = \hat{z}\partial/\partial z + \hat{x} \partial/\partial x$ and the Poisson bracket $\left[f,g\right]_{z,x}= (\pd_z f) (\pd_x g) - (\pd_x f) (\pd_z g)$. A source electric field of magnitude $\hat{\eta}J_{y0}$ is used in Eq. \ref{eqn:psi-2d}, to maintain the equilibrium against resistive dissipation \cite{biskamp1980,biskamp1982,knoll2006POP}. \\ %The above equations are normalized to \Alfvenic units: length $L$ to the system length $L_z$, velocity to \Alfvenic velocity $v_A = B/\sqrt{\mu_0 \rho}$ (free space permeability $\mu_0 = 1$, normalized density $\rho= 1$ is assumed) and time $t$ to \Alfvenic time $t_A = L/v_A$.\\
\begin{figure}[!h]
	\includegraphics[scale=0.4]{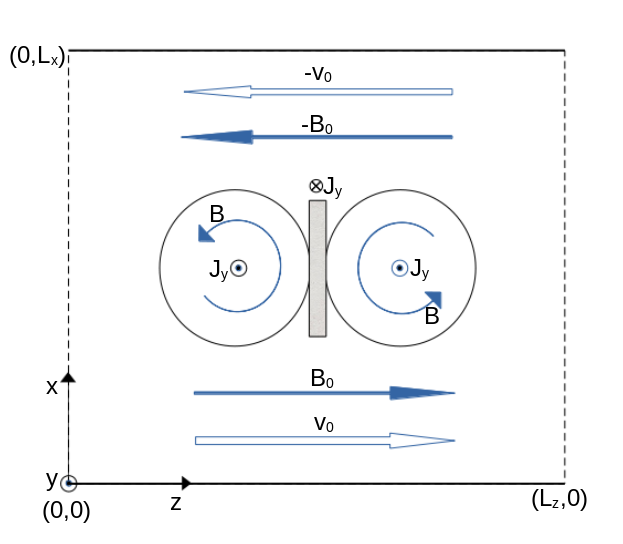}
	\caption{A schematic diagram of the MIC problem showing simulation domain and initial field quantities. {Horizontal solid lines at $x=0,L_x$ location represent conducting boundaries ($B_x = v_x =0$) and vertical dashed lines at $z=0,L_z$ location represent periodic boundaries.}}
	\label{Fig_schematic_plot}
\end{figure}
{As shown in Fig. \ref{Fig_schematic_plot}, Eq. \ref{eqn:omegay-2d}-\ref{eqn:psi-2d} are solved over a 2D uniform grid $0\leq x \leq L_x$ and $0 \leq z \leq L_z$ with $L_z=L_x=2$, and grid sizes $\mathrm{d}z \simeq 10^{-3}$, $\mathrm{d}x \simeq 2\times10^{-3}$ (corresponding $N_z=2048, N_x = 1024)$. The dashed lines at $z=0$ and $L_z$ represents periodic boundaries and the solid lines at $x=0$ and $L_x$ represents conducting walls.} Initial equilibrium profiles are \cite{mahapatra2021},
\begin{equation}
	\left.
	\begin{aligned}
		&J_{y0} = \frac{1-\epsilon^2}{ a_B \left[\cosh \left( \frac{x-L_x/2}{a_B} \right) + \epsilon \cos\left(\frac{z}{a_B} \right) \right]^2} \\
		%& \implies \Psi_{0} = -a_B~ \text{ln}\left[ \cosh \left( \frac{x-L_x/2}{a_B}\right) + \epsilon \cos \left( \frac{z}{a_B}\right) \right]\\
		&\omega_{y0} = \frac{v_0}{a_v \left[\cosh \left(\frac{x-L_x/2}{a_v}\right)  \right]^2} \\
		&\Psi_1 = A \sin(2\pi z/L_z) \sin(\pi x /L_x)
	\end{aligned}
	\right\} \label{eqn:initial_profil}
\end{equation}
Here the parameter $\epsilon$ decides the half-width of current filament $a_I=cosh^{-1}(1+2\epsilon) a_B$ \cite{pritchett2007kinetic} (hence, the total magnetic flux available for reconnection event). {As shown in Fig. \ref{Fig_schematic_plot}, $J_y$ direction in the islands are along positive y-axis and hence, the reconnecting current sheet carries $J_y$ along negative y-axis. For the first case study, a range of $\epsilon$ values are considered in the absence of any shear flow. {For the second case study, $\epsilon=0.2$ is considered and an initial vorticity profile $\omega_{y0}$ (see Eq. \ref{eqn:initial_profil}) is used with a range of shear flow amplitude $v_0$ and shear flow length $a_v$ such that the velocity field are parallel to the asymptotic magnetic field (see Fig. \ref{Fig_schematic_plot}).} The value of $a_B=1/2\pi$ is the thickness of Harris type current sheet when $\epsilon=0$. It decides the periodic domain size ($L_z=4\pi a_B$). {A perturbation in vector potential $\Psi_1$ (see Eq. \ref{eqn:initial_profil}) of initial amplitude $A=10^{-2}$ is added to initiate the instability}. A previous study on shear flow effects on MIC \cite{mahapatra2021} suggests that the peak reconnection time increases with shear flow strength and finite resistive/viscous dissipation of magnetic field and flows could also be responsible for lower reconnection rate. {Hence, as mentioned earlier, a source electric field of magnitude $\hat{\eta} J_{y0}$ is applied (in Eq. \ref{eqn:psi-2d}) to compensate the resistive dissipation for this study.} \\
\indent The reconnection rate is measured as the maximum value of the time varying reconnecting electric field ($E_y$) at the $X$-point, given by the expression $E_y = - \hat{\eta}(J_y -J_{y0})|_{X-point}$. Alternatively, it is also calculated as $E_y(t) = \partial(\psi_O - \psi_X)/\partial t$ where $\psi_O$ and $\psi_X$ are the values of vector potential at O-point and $X$-point, respectively. We have verified that both methods give identical $E_y$ values. However, the later technique cannot be used when the reconnection takes place at multiple and dynamic locations, as observed during the flow driven phase in shear flow affected IC problem \cite{mahapatra2021}. Hence, we have used the former definition for $E_y$ calculation through out this work. For further detailed initial parameters, readers are referred to \cite{mahapatra2021}. \\
\section{Results and Discussion} \label{Results}
\subsection{Variation of island size ($\epsilon$) in absence of shear flow} \label{size-study}
\begin{figure*}[!t]
	\centering
	\includegraphics[scale=0.16]{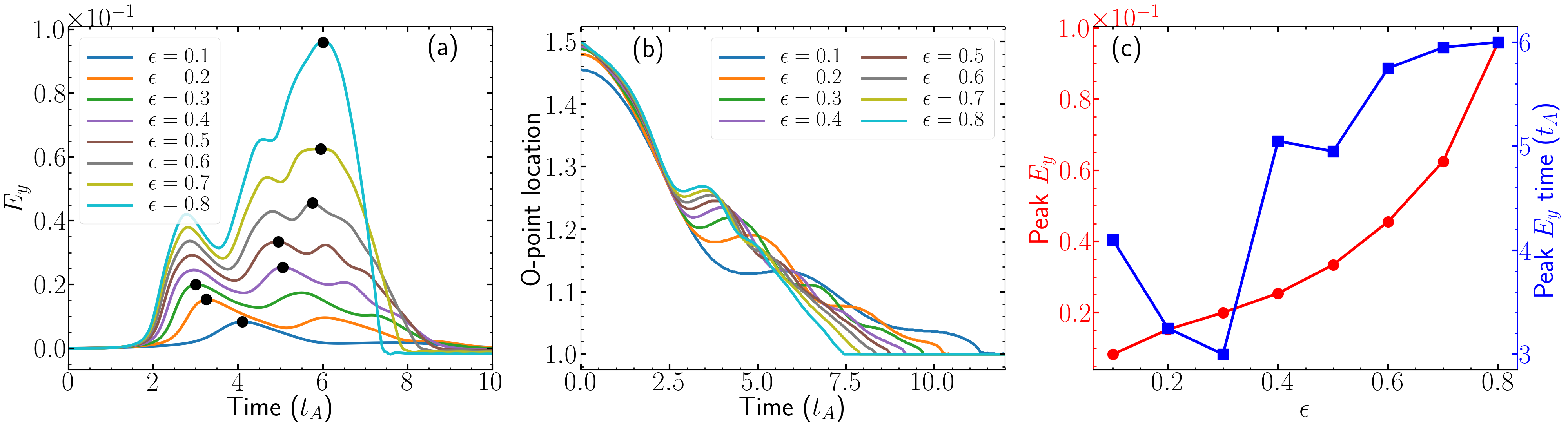}
	\caption{{(a) Time evolution of $E_y$ for different $\epsilon$ values with the black-dots showing their respective maximum reconnection rate {time} location, (b) time function of O-point locations, (c) plot for peak $E_y$ value (red) and peak $E_y$ time (blue) vs. $\epsilon$ values. Here, the reconnection rate is the peak $E_y$ value for each $\epsilon$ case.}}
	\label{Fig-1}
\end{figure*}
In this subsection, a quantitative scaling of reconnection parameters with the magnetic island size $\epsilon$ (in the absence of external shear flow) is presented. The aforementioned parameter $\epsilon$ controls the total magnetic flux enclosed inside the separatrix, and is a proxy to the island size. For this study, 8 runs have been conducted for a range of $\epsilon$ values from $0.1$ to $0.8$ keeping $a_B$ fixed and the initial flow profiles to be zero. For $\epsilon = 0.1 - 0.3$ runs, same simulation parameters are used as mentioned in Section \ref{simulation-details}. For $\epsilon = 0.4 - 0.8$ runs, {due to increasing island width ($a_I$)}, conducting walls are taken away from the islands ($L_x$ is doubled from $L_x=2$ to $L_x=4$ and number of grid points $N_x$ is also doubled to keep the grid size along $x$ i.e. $\mathrm{d}x$ the same) to minimize the boundary effects \cite{pritchett1979coalescence}. This is ensured by maintaining the ratio $L_x/(2a_I) > 6$. \\
\begin{figure}[!h]
	\includegraphics[scale=0.25]{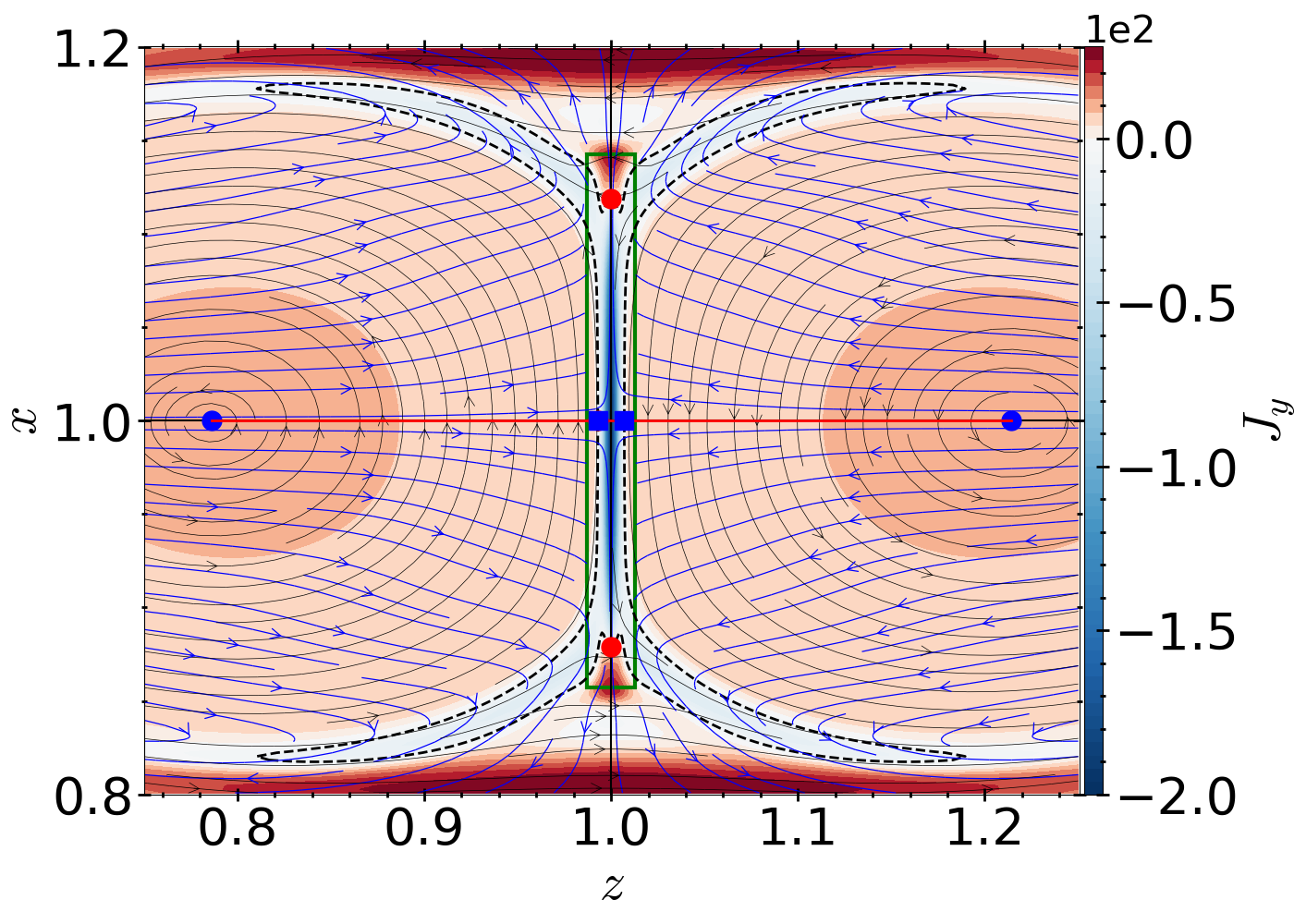}
	\caption{$J_y$ (colormap), magnetic field (black streamline) and flow field (blue streamline) profiles at the peak $E_y$ time $t=3t_A$ for $\epsilon=0.3$. The green rectangle is located at the maximum $\vel \times \B$ values in upstream/downstream region. The black dotted line represents the shape of the current sheet calculated at the location of 1/25th of the $J_y$ value at the $X$-point. The blue circles are the location of O-points. The red colored line, representing the line of approach of magnetic islands, intersects the current sheet boundary (black dotted line) at the blue square dots. The red dot represents the current sheet length edge, at the point of intersection of current sheet boundary and the line along length of current sheet.}
	\label{Fig-2}
\end{figure}
\begin{table}
	\footnotesize
	\caption{Comparison of reconnection parameters for different runs, calculated at the edges of CS where $\vel \times \B$ is maximum. Listed are $\epsilon$, half-width $\delta/2$, half-length $l/2$, upstream velocity $v_{in}$, downstream velocity $v_{out}$, upstream magnetic field $B_{in}$, downstream magnetic field $B_{out}$.}
	\begin{tabular}{c c c c c c c}
		\hline \hline
		\quad$\epsilon$ \quad & \quad $\delta/2$ \quad & \quad $l/2$ \quad & \quad $v_{in}$ \quad &\quad $v_{out}$ \quad & \quad $B_{in}$  \quad & \quad $B_{out}$ \quad \\
		& $\times 10^{-2}$& & $\times 10^{-2}$ & & &  \\
		\hline
		0.1 &	1.07 &	0.06 &	2.97 &	0.20 &	0.28 &	0.07 \\
		0.2 &	1.17 &	0.11 &	3.03 &	0.41 &	0.53 &	0.10 \\
		0.3 &	1.27 &	0.14 &	3.04 &	0.55 &	0.70 &	0.18 \\
		0.4 &	1.86 &	0.11 &	3.99 &	0.65 &	0.69 &	0.22 \\
		0.5 &	0.88 &	0.14 &	3.53 &	0.76 &	0.98 &	0.36 \\
		0.6 &	0.68 &	0.11 &	4.22 &	0.87 &	1.14 &	0.43 \\
		0.7 &	0.59 &	0.11 &	4.65 &	1.09 &	1.41 &	0.11 \\
		0.8 &	0.49 &	0.11 &	5.56 &	1.44 &	1.84 &	0.44 \\
		\hline \hline
	\end{tabular}
	\label{table1}
\end{table}
\begin{figure*}[!t]
	\includegraphics[scale=0.177]{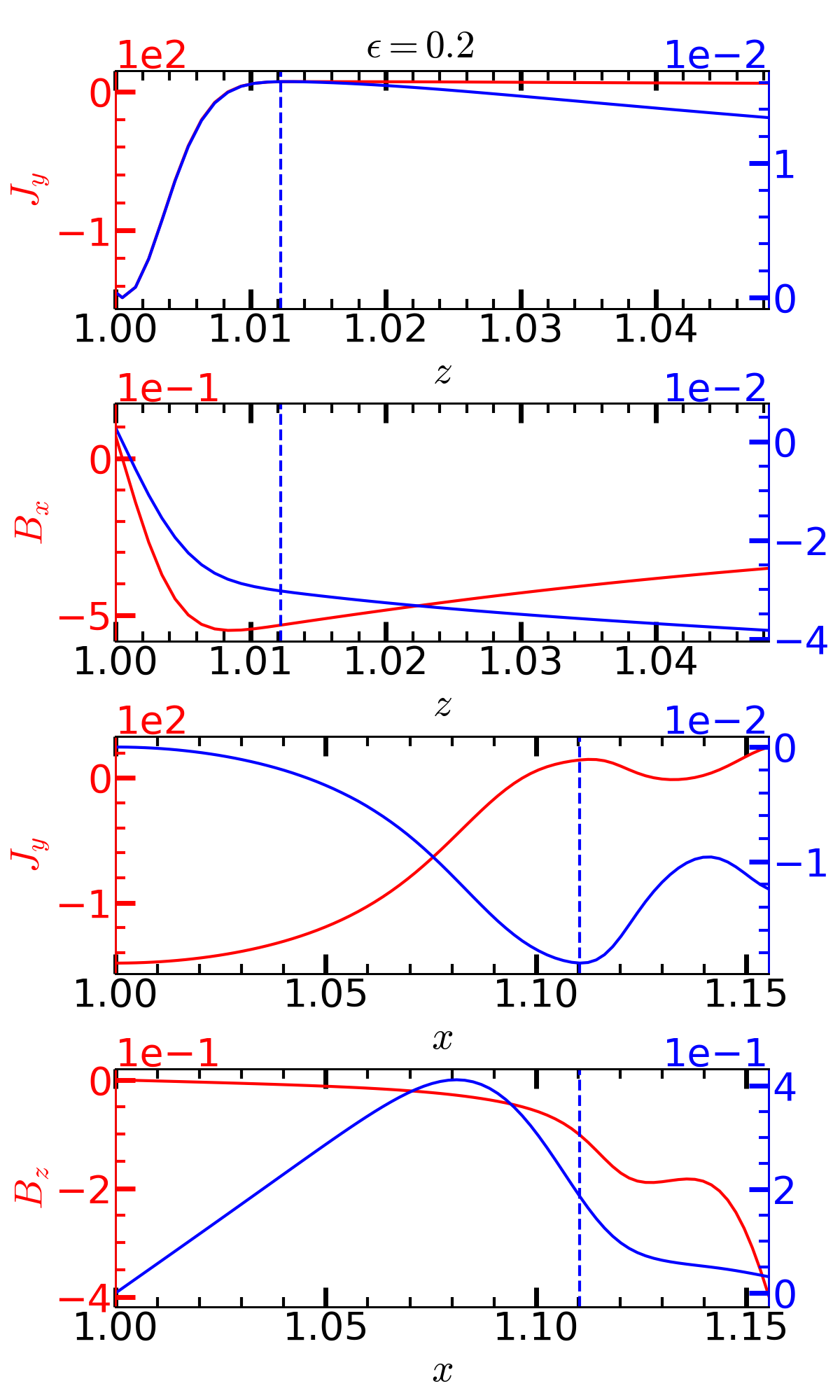}
	\includegraphics[scale=0.177]{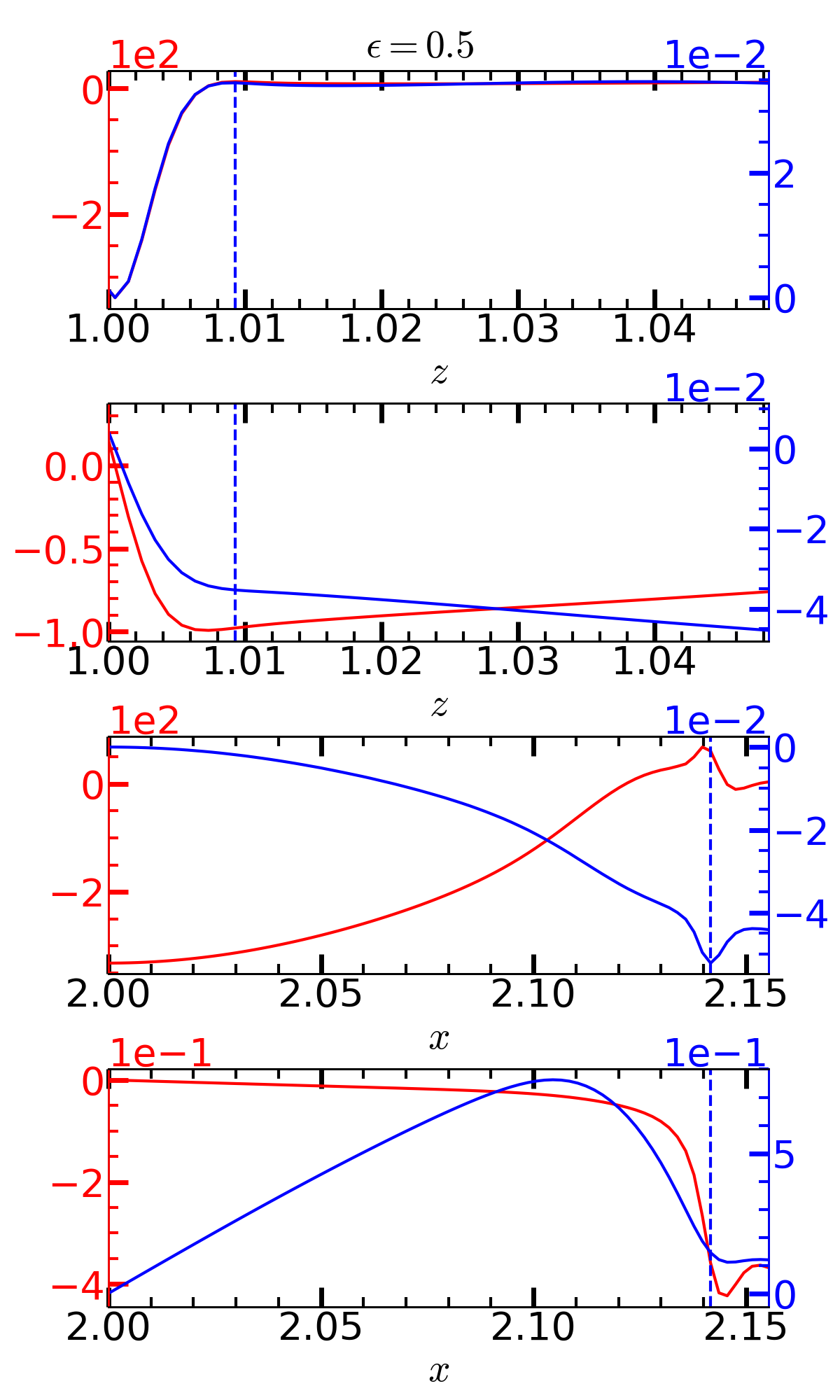}
	\includegraphics[scale=0.177]{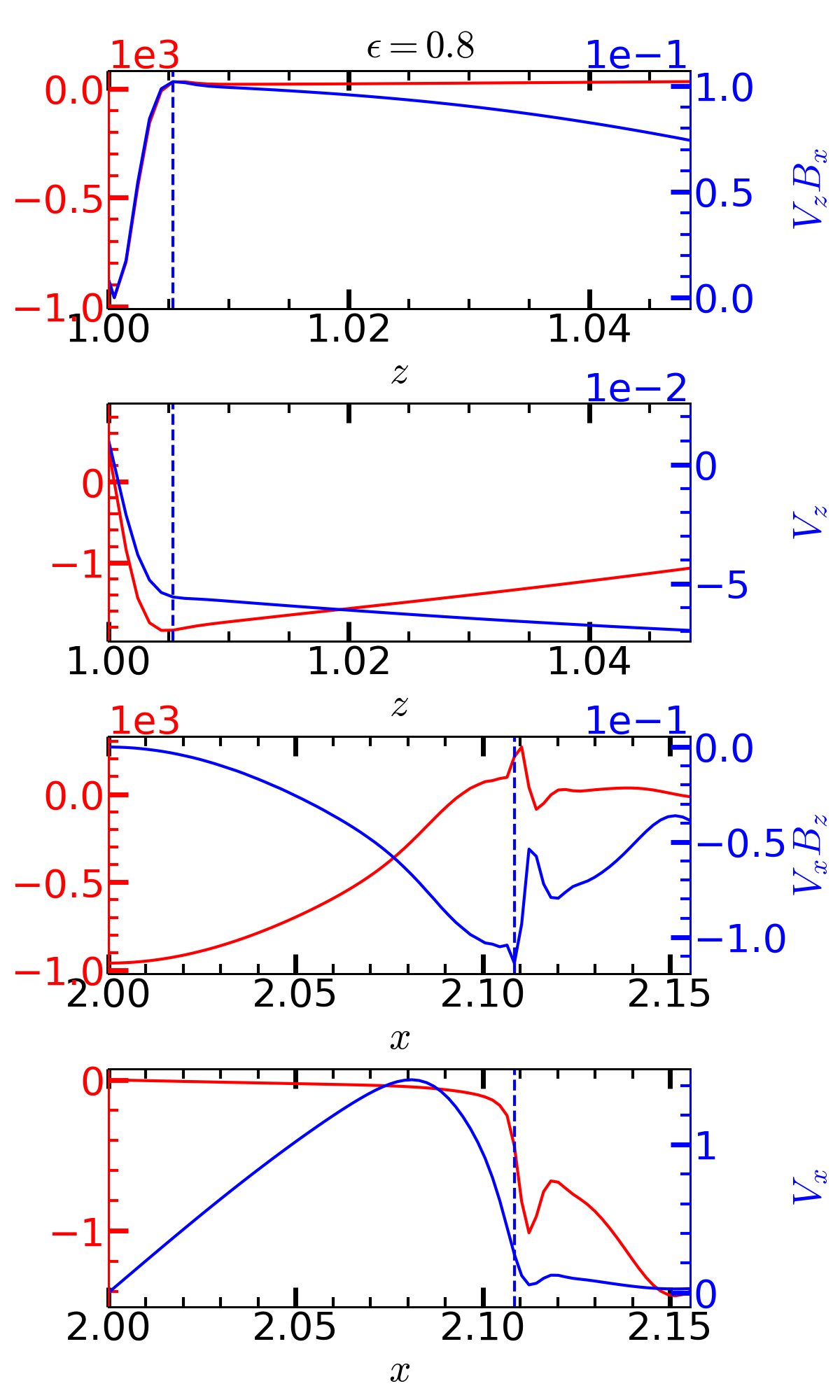}	
	\caption{$J_y$, $v_z B_x$, $v_z$, $B_x$ are plotted along $z$-axis (at $x=L_x/2$) in the first two rows of each column,  and $J_y$, $v_x B_z$, $v_x$, $B_z$ are plotted along $x$-axis (at $z=L_z/2$) on bottom two rows of each column for $\epsilon = 0.2$ (left column), $\epsilon=0.5$ (middle column), and $\epsilon=0.8$ (right column) at the time of peak $E_y$. In the plots of first two rows, the blue dotted line shows the location of current sheet half-width ($\delta/2$), where $v_z B_x$ attains maximum. Likewise, in the last two rows, the blue dotted line shows location of current sheet half-length ($l/2$) where $v_x B_z$ attains maximum. }
	\label{Fig-3}
\end{figure*}
\indent The time varying $E_y$ line-plots for all the runs are shown in Fig. \ref{Fig-1}a. Multiple humps in each $E_y$ plot are observed due to {the pile-up of magnetic flux in the upstream region (where the plasma flows into the CS) that causes island sloshing}, a well studied phenomena in MIC problem \cite{knoll2006POP, mahapatra2021, pritchett1979coalescence, biskamp1982}. The peak $E_y$ values, denoted as black dots in Fig. \ref{Fig-1}a, and their corresponding time locations are plotted as a function of $\epsilon$ value in Fig. \ref{Fig-1}c. For smaller $\epsilon$ values (0.1 to 0.3), most of the flux reconnects before the development of sufficient upstream magnetic pressure to slow down the coalescence. Hence, for this range of $\epsilon$, the size of first hump is larger than the second. Eventually, remaining flux reconnects during the second hump and coalescence is completed. With increasing $\epsilon$, the size of the first hump gradually increases where as that of second hump increases by many fold. Moreover, with increasing strength of the Lorentz force between the islands, peak value of $E_y$ in first hump is attained at the earlier times. Simultaneously, the magnetic pressure also builds-up on both the sides of CS at earlier times for larger islands. This is evident as the dip in $E_y$ value after first hump in Fig. \ref{Fig-1}a and flattening of O-point tracing lines in Fig. \ref{Fig-1}b at earlier times as $\epsilon$ increases.  In case of larger islands ($\epsilon \geq 0.4$), at the end of first hump, a large fraction of magnetic flux still remains inside the islands to be reconnected, resulting in the second hump to be much bigger in size and the coalescence continues for a longer time (compared to first hump). A jump in the peak $E_y$ time between $\epsilon=0.3 $ and 0.4 case (blue line-plot in Fig. \ref{Fig-1}c) is observed as the location of peak $E_y$ value is shifted from first hump to second hump during this transition. {Furthermore, in Fig. \ref{Fig-1}b, from time evolution of O-point locations, the complete coalescence time  for $\epsilon = 0.1, 0.2, 0.3, 0.4, 0.5, 0.6, 0.7,$ and $0.8$} are found to be $t=11.5t_A, 10.25t_A, 9.75t_A, 9.2t_A, 8.75t_A, 8.4t_A, 8t_A$ and $7.5t_A$. This suggests the larger current filaments take lesser time to complete the coalescence compared to the smaller ones.\\
\begin{table*}
	%\footnotesize
	\centering
	\caption{Comparison of reconnection parameters for each run. Listed are $\epsilon$, current sheet half-width $\delta/2$, current sheet half-length $l/2$, upstream velocity $v_{in}$, downstream velocity $v_{out}$, upstream B-field $B_{in}$, downstream B-field $B_{out}$, for mass conservation check $v_{in}l/v_{out} \delta$, unnormalized reconnection rate $E_y$,  $v_{in}B_{in}$ at the edge of current sheet, aspect ratio of current sheet $\delta/l$, normalized rate $\tilde{E_y}=E_y/(v_A^{in}B_{in})$ and normalized upstream magnetic field $B^* = B_{in}/B_{x0}$.}
	\begin{tabular}{c c c c c c c c c c c c c c} 
		\hline \hline
		$\epsilon$ \quad &  $\delta/2$ \quad & $l/2$ \quad & $v_{in}$ \quad & ~$v_{out}$ ~\quad & $B_{in}$  \quad & ~ $B_{out}$~ \quad & $\frac{v_{in}l}{v_{out} \delta}$ \quad & $\frac{B^2_{in}}{v^2_{out}}$ \quad & $E_y$ \quad & $v_{in}B_{in}$ \quad &  $\delta/l$ \quad & $\tilde{E_y}$ \quad & $B^*$\\ [0.5ex] 
		&$\times 10^{-3}$ & $\times 10^{-2}$ & $\times 10^{-2}$ & & & $\times 10^{-2}$ & & & $\times 10^{-2}$ &  $\times 10^{-2}$ &  & $\times 10^{-2}$ & \\
		\hline
		0.1 & 6.79 & 5.59  & 2.64 & 0.18 & 0.29 & 4.15 & 1.20 & 2.50 & 0.83 & 0.75  & 0.12 & 10.1 & 3.68 \\ 		
		0.2 & 7.06 & 9.32  & 2.79 & 0.37 & 0.55 & 4.25 & 0.99 & 2.16 & 1.53 & 1.53  & 0.08 & 5.09 & 3.04 \\		
		0.3 & 7.09 & 11.91 & 2.79 & 0.50 & 0.72 & 4.13 & 0.93 & 2.03 & 2.0  & 2.02  & 0.06 & 3.84 & 2.5  \\		
		0.4 & 5.88 & 9.22  & 3.36 & 0.53 & 0.78 & 5.01 & 1.00 & 2.15 & 2.54 & 2.61  & 0.06 & 4.22 & 1.89 \\		
		0.5 & 5.76 & 11.78 & 3.31 & 0.69 & 0.99 & 4.65 & 0.99 & 2.07 & 3.34 & 3.27  & 0.06 & 3.43 & 1.79 \\		
		0.6 & 4.83 & 9.23  & 3.95 & 0.78 & 1.14 & 5.72 & 0.97 & 2.14 & 4.56 & 4.51  & 0.05 & 3.5  & 1.58  \\		
		0.7 & 4.42 & 9.64  & 4.53 & 0.97 & 1.42 & 5.92 & 1.02 & 2.12 & 6.25 & 6.41  & 0.05 & 3.12 & 1.49 \\		
		0.8 & 3.72 & 9.11  & 5.36 & 1.31 & 1.84 & 6.74 & 1.00 & 1.97 & 9.6  & 9.88  & 0.04 & 2.83 & 1.41 \\
		\hline \hline
	\end{tabular} 
	\label{table2}
\end{table*}
\begin{figure*}[t]
	\includegraphics[scale=0.42]{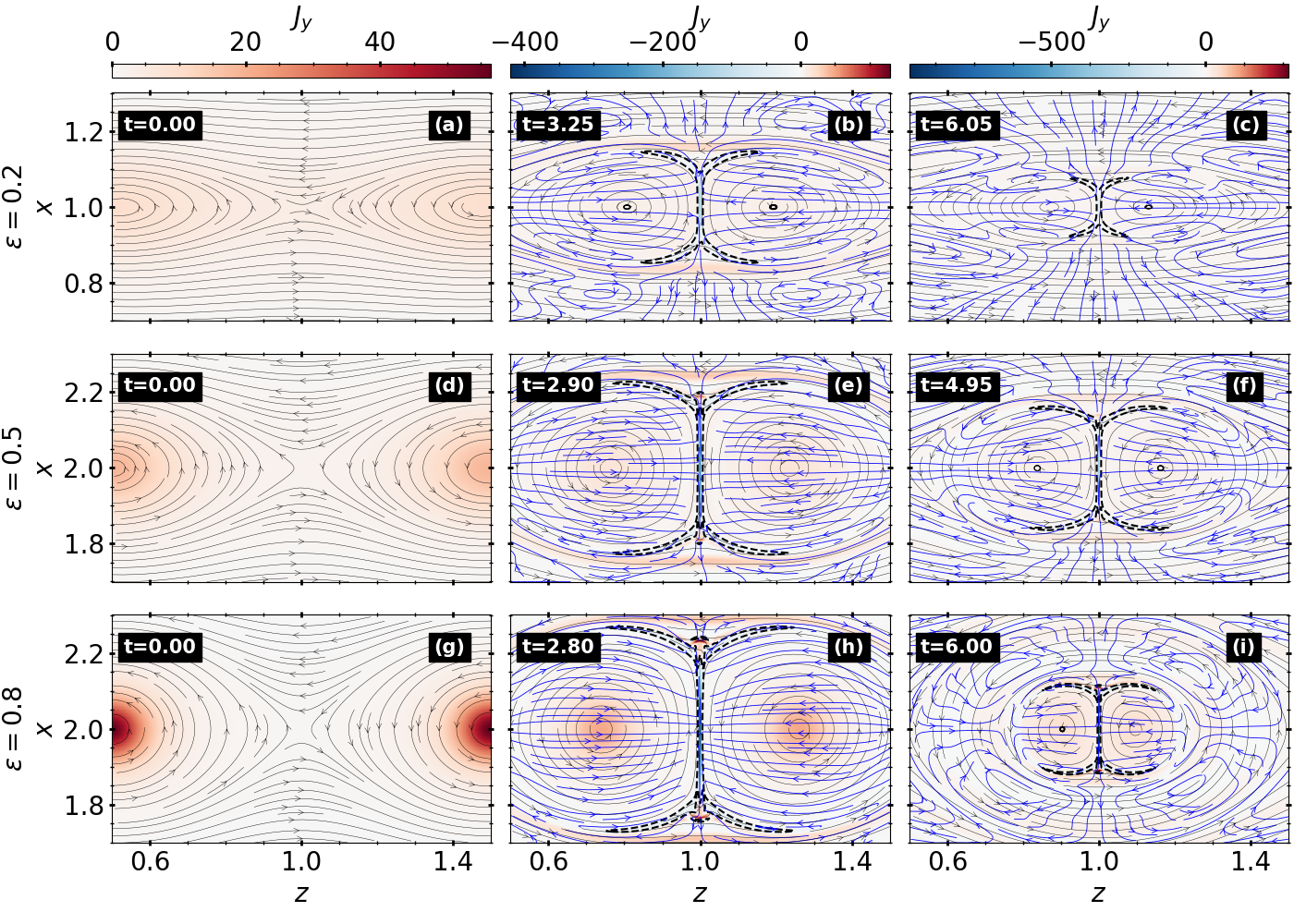}
	\caption{Time evolution of $J_y$ (colormap), $\B$ (black streamlines) and $\vel$ (blue streamlines) spacial profiles for $\epsilon$ = 0.2 (a, b, and c subplot), 0.5 (d, e, and f subplot), and 0.8 (g, h, and i subplot). The spatial profiles in the first column (subfig. a, d, and g) is at the time $t=0$, second column (subfig. b, e, and h) is at the peak $E_y$ time in the first hump and third column (subfig. c, f, and i) is at the peak $E_y$ time in the second hump as shown in Fig. \ref{Fig-1}a. The length of the reconnecting CS at peak $E_y$ time for $\epsilon=0.2$ (b), $\epsilon=0.5$ (f) and $\epsilon=0.8$ (i) are almost equal, as evident in Table II.}
	\label{Fig-4}
\end{figure*}
\indent To further investigate the CS properties, various quantities such as $J_y$, velocity and magnetic field components are calculated in both upstream and downstream region for each $\epsilon$ run. A zoomed plot of a fully developed reconnecting sheet for $\epsilon=0.3$ is shown in Fig. \ref{Fig-2}. First, the shape of the diffusion region is calculated at the location of maximum $\vel \times \B$ value \cite{knoll2006POP, knoll2006PRL}, represented by the green rectangle in Fig. \ref{Fig-2}. {Here, the region where plasma flows into the CS is known as upstream region. Homologous to this, the region where the plasma comes out of the CS is known as downstream region.} In Table \ref{table1}, we have listed CS half-thickness ($\delta/2$, distance from the $X$-point to the location of maximum $v_z B_x$), half-length ($l/2$, distance from $X$-point to the location of maximum $v_x B_z$) along with upstream/downstream magnetic and velocity field components, for each $\epsilon$ value. \\
\indent {In Fig.\ref{Fig-3}, above quantities are plotted across and along the CS length for three different $\epsilon$ values at their respective peak $E_y$ time. The distance $\delta/2$ and $l/2$ are marked as vertical blue dotted line at the location of maximum values of $v_z B_x$ and $v_x B_z$, respectively. From Fig. \ref{Fig-3} and Table \ref{table1}, one can notice a gradual decrease in $\delta/2$ with increasing $\epsilon$ value. For large $\epsilon$ cases, full-width of CS becomes more close to the Sweet-Parker \cite{parker1957sweet} scaling $\delta \sim \sqrt{\hat{\eta}}$ (=0.01 for $\hat{\eta}=10^{-4}$). Notably, in Fig. \ref{Fig-3}, the upstream $|v_z|$ and $|B_x|$ attains maximum at $\delta$ location. However, in the downstream region, location of maximum $|v_x|$ and $v_x B_z$ are different. Using $v_x$ at half-length edge, the values of mass conservation ratio $|v_{in}l/v_{out} \delta|$ are 1.434, 1.52, 2.24, 1.564, 3.888, 4.430, 1.459, 4.713. Therefore the above method of CS length and downstream values calculation is less accurate.} \\
\indent In the \textit{second} method, the current sheet half-width $\delta/2$ and half-length $l/2$ are calculated at a distance of $1/25^{th}$ from central peak $J_y$ value, at the time of peak $E_y$. In Ref. \cite{murtas2021coalescence} (see Section III.B), a similar method is reported for $\delta$ and $l$ calculation. The upstream/downstream field components ($v_{in}, B_{in}, v_{out}, B_{out}$), half-thickness ($\delta/2$), half-length ($l/2$) and other parameters calculated by this technique are listed in Table \ref{table2}. The $v_{in}$ and $B_{in}$ are the respective $z$ and $x$ components at the half-width edge of the current sheet (at the blue square marks in the Fig. \ref{Fig-2}). Likewise, $v_{out}$ and $B_{out}$ are the respective $x$ and $z$ components at the half-length edge (at the red circular dots in the Fig. \ref{Fig-2}). Corroborating the Sweet-Parker \cite{parker1957sweet} model, the quantity $v_{in} l / v_{out} \delta \sim 1$ verifies the mass conservation principle.
This also shows $v_{in}/v_{out} = \delta/l$. Likewise, $B^2_{in}/v^2_{out} \sim 2$ verifies the energy conservation principle in the vicinity of reconnecting sheet. One can notice, as island size increases, the length of the current sheet remains nearly constant but half-thickness decreases by $50\%$. This in turn increases $v_{in}$ as $v_{in} \delta/2=\eta$ (unnormalized resistivity $\eta = 2 \times 10^{-4}$) \cite{sweet1958,parker1957sweet}. Larger islands contain more magnetic flux resulting in an increase in $B_{in}$ with $\epsilon$. However, $B_{out}$ value remains nearly constant. The values of $v_{in}B_{in}$ just outside the current sheet are equal to peak $E_y$ values. However, similar to the results of Ref. \cite{knoll2006POP}, the values of $\delta/l$ are different from peak $E_y$. This anomaly in the current sheet aspect ratio for coalescence problem is also observed while investigating the peak $E_y$ at different $\hat{\eta}$ \cite{mahapatra2021,knoll2006POP}. The amount of magnetic pressure build up in the upstream regime for different size islands can be calculated from the ratio of $B_{in}$ to initial maximum magnetic field value along the line joining the O-points $B_{x0}$ \cite{karimabadi2011flux}. With increasing $\epsilon$ value, this ratio comes out to be 3.68, 3.04, 2.5, 1.89, 1.79, 1.58, 1.49, and 1.41. This indicates the flux pile up decreases with increase in island size, possibly due to the increasing strength of the Lorentz force. {The unnormalized reconnection rate (peak $E_y$), even-though it increases monotonically, the normalized rate $\tilde{E_y} = E_y/(B_{in}v_A^{in})$ ($v_A^{in} = B_{in}$) weakly depends on island size, which is consistent with the results of Ref. \cite{karimabadi2011flux}.} \\
\indent Figure \ref{Fig-4} shows the time evolution of different sized islands and their $J_y$, $\B$ and $\vel$ profiles. We plot a comparison of the island size and current sheet size at initial times (shown in subplots of first column), first peak and second peak in the time varying $E_y$ plots (shown in second and third column subplots, respectively). As time progress, the island size decreases for any fixed $\epsilon$ value and hence, the CS length \cite{Jonathan2015}. We observe the CS length for $\epsilon = 0.2, 0.5, 0.8$ at time t = 3.25$t_A$, 4.95$t_A$ and 6.0$t_A$ (corresponds to peak $E_y$ time) are of same value: suggesting that the CS length is independent of island size. In conclusion, the current sheet thickness and upstream velocity are found to be a function of the magnetic island size, at a fixed Lundquist number $S$ ($= 1/\hat{\eta}$). The unnormalized peak $E_y$ increases with island size due to an increase in upstream magnetic field, although normalized peak $E_y$ remains roughly constant. The upstream magnetic pressure is found to be decreasing with increasing island size. Unlike the Harris sheet reconnection and Sweet-Parker model, the rate of reconnection (peak $E_y$) in MIC problem is different from $v_{in}/v_{out}$ ratio and CS aspect ratio $\delta/l$ (see column 10 and 12 in Table \ref{table2}).

\subsection{Influence of shear flow parameters} \label{shear-flow-study}
In this subsection, we discuss the shear flow effect on different reconnection parameters in the vicinity of diffusion region. The in-plane shear flow is initialized in the form of a vorticity sheet ($\omega_{y0}$ profile in Eq. \ref{eqn:initial_profil}) along with a double island system, such that the velocity field is parallel to the asymptotic magnetic field. Here our approach is similar to that reported in Ref \cite{mahapatra2021}. To study the effect of shear flows applicable to various physical systems, a range of shear flow amplitudes ($v_0$) is chosen from $0.1v_A$ (sub-Alfv\'enic) to $1.4v_A$ (super-Alfv\'enic) values along with four different length scales ($a_v = 2a_B, a_B, 0.5a_B, 0.25a_B$). {Here it is important to highlight that, unlike the externally driven current profiles, the velocity field decays over time due to finite viscosity.} {For our set of parameters and domain size, in the absence of magnetic islands ($\epsilon =0$), shear flows of $a_v \geq a_B$ are susceptible to MHD-KHI modes}; the fastest MHD-KHI mode $m$ satisfies the condition $m=L_z/2\pi a_v > 1$. However, shear flow having $a_v = 2a_B$ does not support any unstable KHI mode ($m < 1$). Stability analysis of MHD-KHI modes for above mentioned flow profile has been discussed in Section IV of Ref \cite{mahapatra2021}.\\
\begin{figure}[!h]
\includegraphics[scale=0.23]{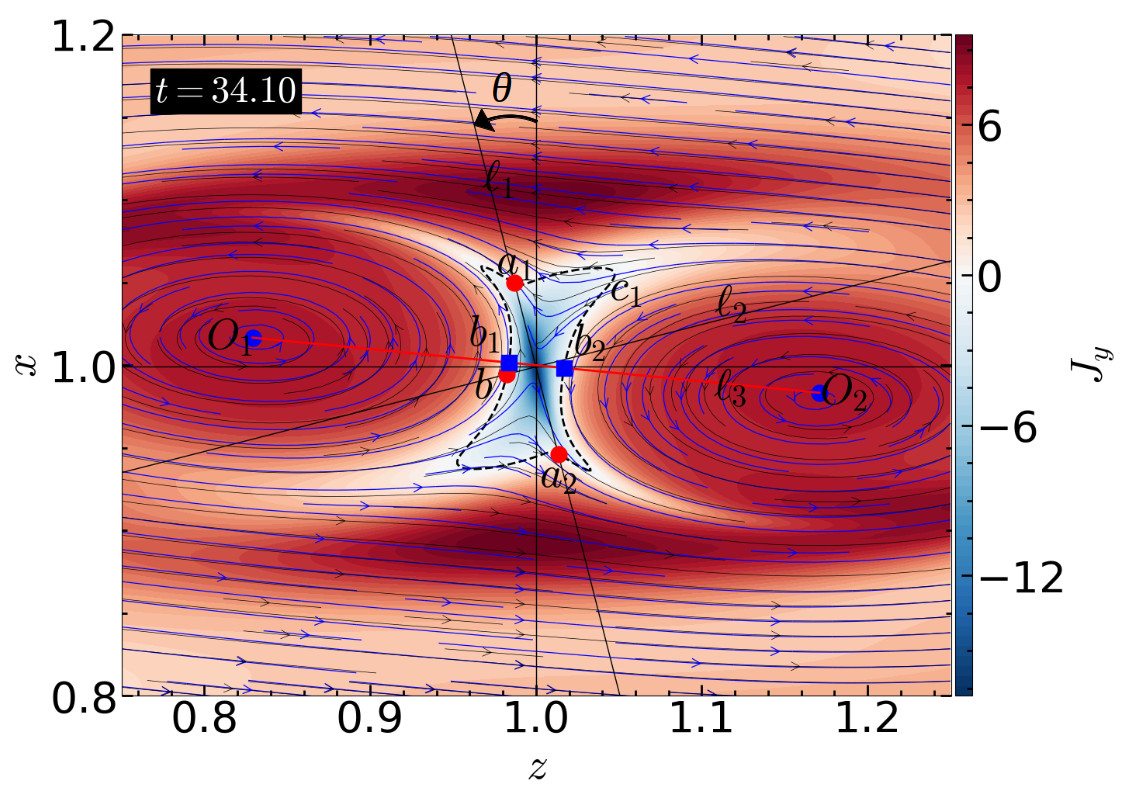}%{Fig_jy_av-1aB_v01.2e0}
\caption{A zoomed plot of $J_y$ (colormap), $\B$ (black streamlines), and $\vel$ (blue streamlines) for $a_v=a_B$ and $v_0 = 1.2 v_A$ is plotted at the peak $E_y$ time. The black dotted line ($c_1$) shows the contour of 1/25th of $J_y|_{X-point}$, encircles the diffusion region. The lines $\ell_1$ and $\ell_2$ representing the directions along and across the current sheet, intersect $c_1$ at points $a_1$ and $a_2$ and $b$ respectively. Likewise, two O-points $O_1, O_2$ connected through the line $\ell_3$, intersects $c_1$ at points $b_1$ and $b_2$. The distance $b_1 b_2$ and $a_1 a_2$ represents current sheet thickness and length respectively.}
\label{Fig-flow-zoomJy}
\end{figure} 
\begin{figure}[!h]
	\includegraphics[scale=0.25]{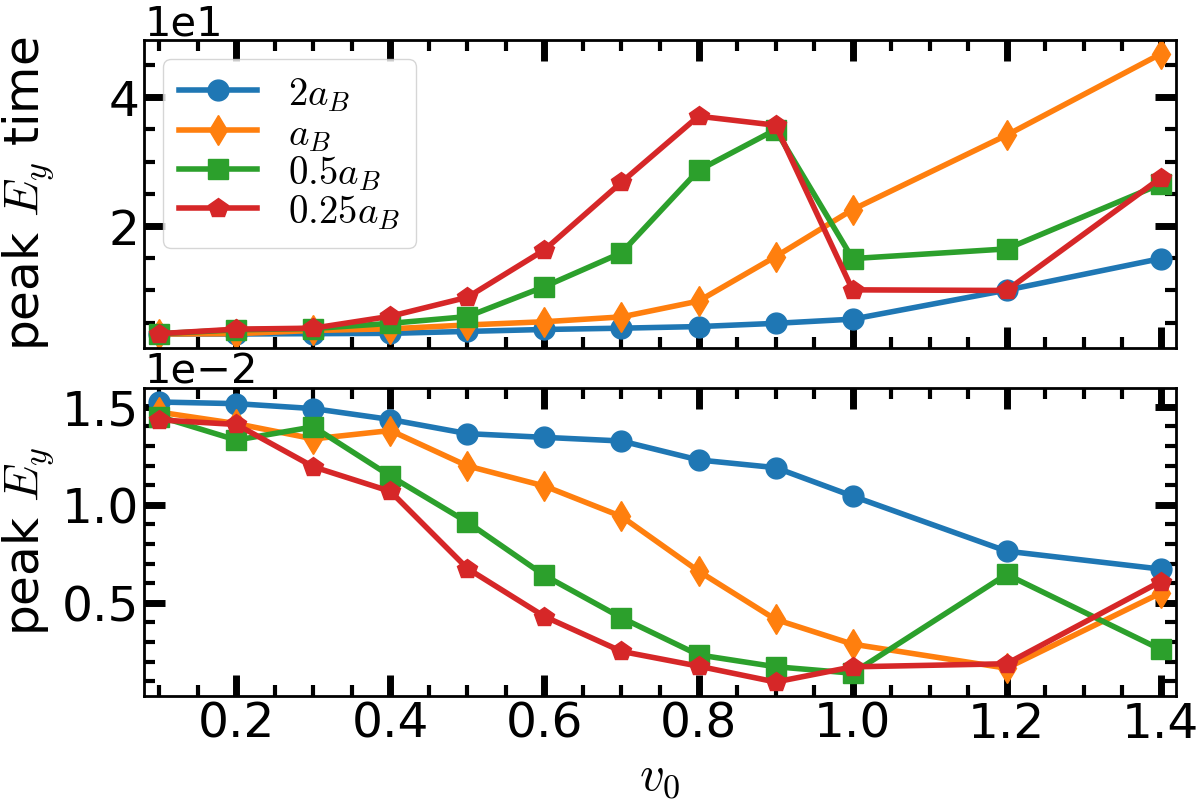}
	\caption{Peak $E_y$ and peak $E_y$ time as a function of $v_0$ for different $a_v$ values.}
	\label{Fig-recRate-ScanV0}
\end{figure}
\begin{figure*}[!t]
	\includegraphics[scale=0.3]{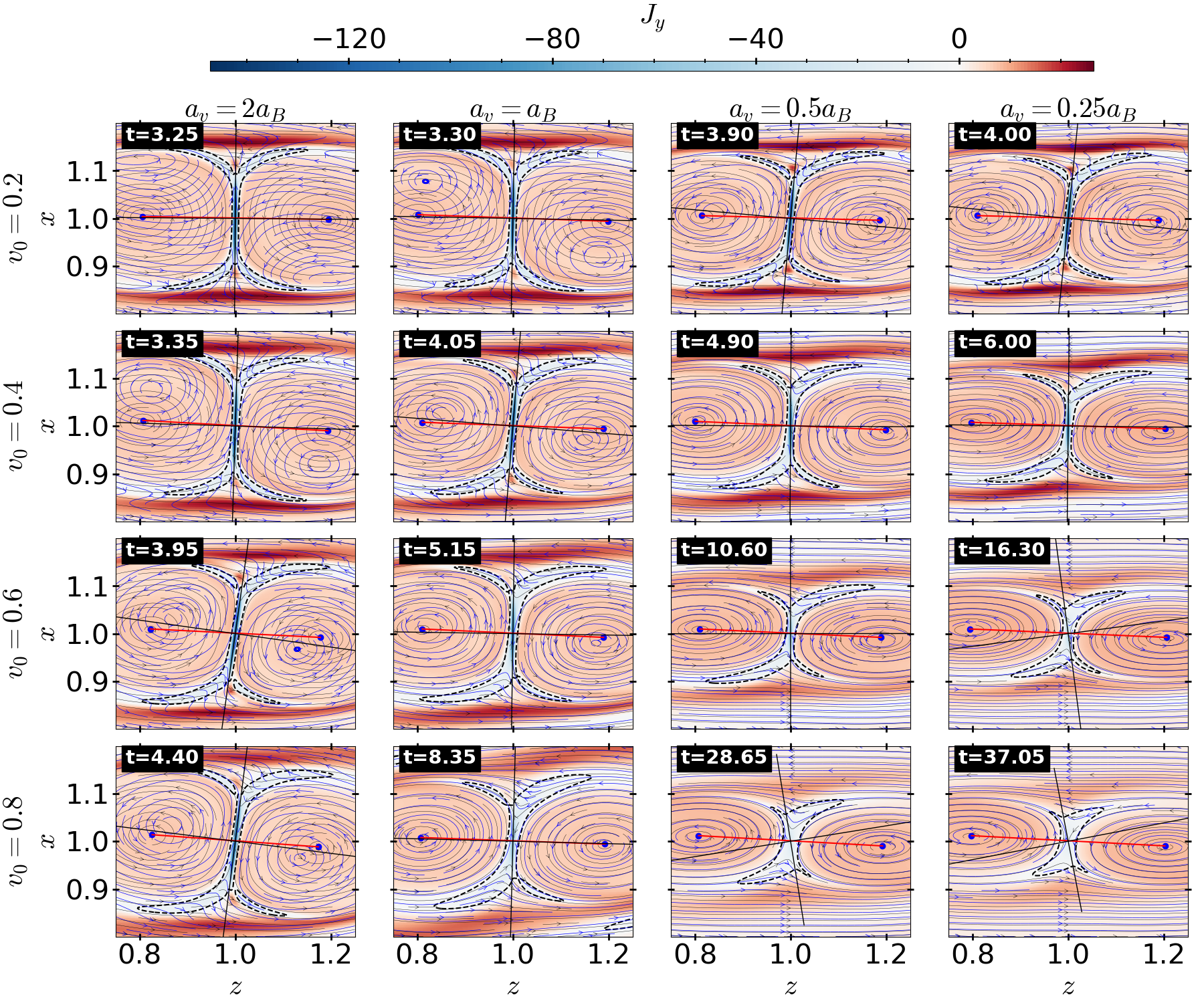}
	\caption{Zoomed profiles of $J_y$ (colormap), $\B$ (black streamlines), and $\vel$ (blue streamlines) close to the reconnecting current sheet for four $a_v$ values (column wise) and four values of $v_0$ at their respective peak $E_y$ time (same plot as Fig. \ref{Fig-flow-zoomJy}). The $x$-limit (vertical) and $z$-limit (horizontal) are [0.8,1.2] and [0.75, 1.25], respectively.}
	\label{Fig-flow-groupJy}
\end{figure*}
\indent By introducing the in-plane shear flow to the island system, we observe a complete change in the upstream/downstream flow pattern during the coalescence. This is evident from Fig. \ref{Fig-flow-zoomJy} showing a zoomed plot of different profiles close to the diffusion region for the flow parameter $a_v=a_B$ and $v_0=1.2v_A$. A group of similar plots for different shear flow parameters are shown in Fig. \ref{Fig-flow-groupJy}. As is evident in all these plots, depending on the plasma flow pattern, the whole region close to the CS can be divided into two separate regions: the asymptotic flow region far from the islands along the y-direction (we name it as primary flow having amplitude $v_0$) and the plasma circulation flow region inside the magnetic islands known as secondary flow induced due to the primary flow. In Fig. \ref{Fig-flow-zoomJy}, the region $x<0.9$ and $x>1.1$ is dominated by the primary flow where as the secondary flow is dominant in a region $0.9 \leq x \leq 1.1$. Unlike Fig. \ref{Fig-2}, the velocity streamlines due to this secondary flow are concentric to magnetic field lines and generate a secondary shear flow on both sides of the reconnecting current sheet, having its magnitude proportional to $v_0$. Likewise, the total coalescence process over time can be divided into two phases, {\it{first}} is the flow dominant phase and {\it{second}} is the coalescence phase. {During the flow dominant phase, in the absence of MHD-KH modes ($a_v =2 a_B$), the overall shape of the magnetic islands remains undisturbed by the primary shear flow, the secondary circulations stabilize them against the Lorentz force of attraction and prevents coalescence. For $a_v \leq a_B$ cases (with unstable MHD-KH modes), during the first phase, the primary flow severely distorts the island shape and displaces them from their original location. The strong shear flow convect the islands vigorously and whenever they come close to each other a fraction of the total magnetic flux gets reconnected. Hence the reconnection becomes bursty. We find this to be analogous to vortex-induced-reconnection as reported in literature, though there are no KH-vortices. After a sufficient time, the flow dissipates down and the coalescence phase begins where the islands regain their shape. The regained size of magnetic islands are smaller than their initial size due to the bursty reconnection during the first phase. Hence, the upstream magnetic flux $B^{up}_x$ (shown in Fig. \ref{Fig-flow-parameter-scaling}) and the peak reconnection rate (shown in Fig. \ref{Fig-recRate-ScanV0}) decreases with an increase of shear flow strength.} Moreover, for the stronger shear flow cases, the flow dominant phase lasts for longer time and hence the peak reconnection time (pean $E_y$ time) increases as shown in Fig. \ref{Fig-recRate-ScanV0}. \\
\indent In Fig. \ref{Fig-flow-zoomJy}, at the peak $E_y$ time for $a_v = a_B$ and $v_0=1.2.v_A$, we observe that the shear flows have displaced the O-points ($O_1, O_2$) from $x=L_x/2$ line; the line $\ell_3$ joining them is inclined towards $z$-axis in clockwise direction. The diffusion region (enclosed by the contour $c_1$) is also inclined towards the clockwise direction with respect to vertical $x$-axis (approximately perpendicular to $\ell_3$). However, the current sheet's inclination is in opposite direction to that of $c_1$. The same plot for some other $v_0$ and $a_v$ values are plotted in Fig. \ref{Fig-flow-groupJy}. One can observe that for weak shear flow cases ($a_b = 2a_B$ case or $v_0 \simeq 0.1v_A$ case for $a_v \leq a_B$), the shear flow streamlines are not coinciding with that of magnetic field at their peak $E_y$ time. However with increasing shear flow strength, the peak $E_y$ time increases and the velocity, magnetic field streamlines become concentric. This suggests that the concentric vorticity patches do not form immediately after the flow initialization. Moreover, for $a_v < a_B$ cases, when the $v_0$ is higher, the inclination angle of the current sheet becomes different from the diffusion region. For these cases the current sheet tilt is such that the pressure exerted by the shear flow on it is minimum \cite{cassak2011}. \\
\begin{table}[!h]
	\centering
	\caption{Comparison of current sheet inclination angle ($\theta$, in degree, see Fig. \ref{Fig-flow-zoomJy}) for different values of $v_0$ and $a_v$. With reference to positive $x$-axis, $\theta$ in anticlockwise direction is taken positive and vice-versa.}
	\begin{tabular}{c c c c c}
		\hline \hline
		$v_0$ \quad & \quad $a_v=0.25a_B$ \quad & \quad $a_v=0.5a_B$ \quad & \quad $a_v=a_B$ \quad & \quad $a_v=2a_B$ \quad \\
		\hline
		0.1 & -0.86  &-0.86  &-0.57  & -0.16 \\
		0.2 & -5.73  &-5.16  &-1.03  & -0.46 \\
		0.3 &  3.44  &-1.15  &-5.73  & -1.43 \\
		0.4 & -0.57  &-0.34  &-4.58  & -1.72 \\
		0.5 &  4.58  & 1.03  &-2.01  & -5.16 \\  
		0.6 &  7.56  & 0.06  &-0.86  & -8.02 \\
		0.7 &  8.59  & 6.3   &-0.86  & -7.45 \\
		0.8 &  10.89 & 9.17  &-1.45  & -7.22 \\
		0.9 &  18.91 & 10.89 & 0.91  & -6.42 \\
		1.0 &  -     & 14.32 & -0.31 & -6.59 \\
		1.2 &  -     & -     & 14.32 & -8.02 \\
		1.4 &  -1.72 & 2.72  & -     & -3.44 \\
		\hline \hline
	\end{tabular}
	\label{table3}
\end{table}
\begin{figure}[!h]
	\centering
	\includegraphics[scale=0.25]{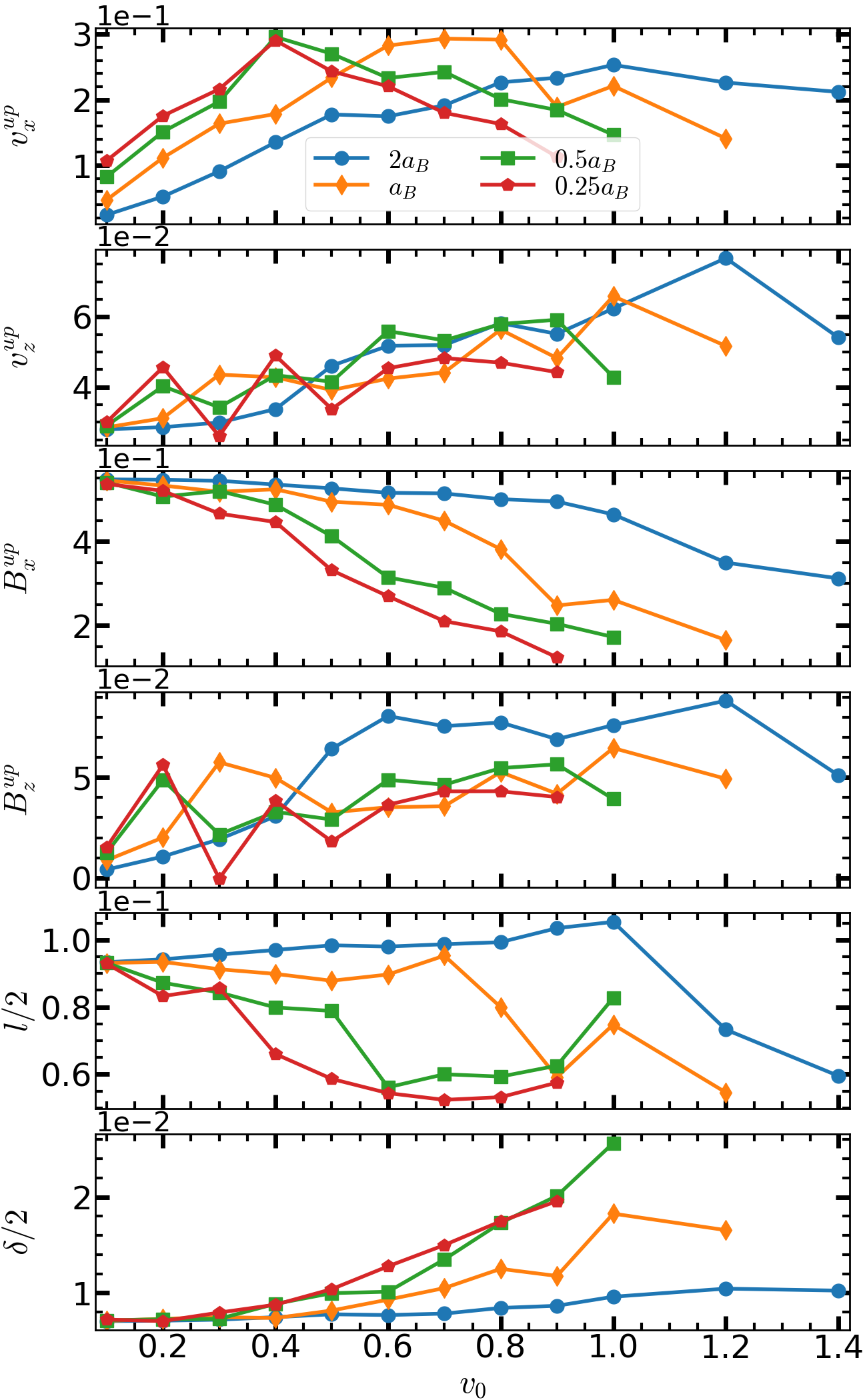}
	\caption{Plot of upstream magnetic field ($B^{up}_x$ and $B^{up}_z$), velocity field ($v^{up}_x$ and $v^{up}_z$) components, current sheet half-width ($\delta/2$) and half-length ($l/2$) for different values of $v_0$ at the peak $E_y$ time. Colored line plots represent different $a_v$ values: $a_v = 2a_B$ (blue), $a_B$ (orange), $0.5a_B$ (green), and $0.25a_B$ (red).}
	\label{Fig-flow-parameter-scaling}
\end{figure}
\indent {One of the methods for the CS inclination angle calculation (see Ref. \cite{cassak2011} for Harris sheet set-up) uses a Fourier padding technique to locate the X-point followed by the Hessian calculation in a doubly periodic domain. However, Fourier padding cannot be used along x-directional conducting boundaries and hence the Hessian calculations will be inaccurate}. Alternatively, we calculate the angle by fitting a straight line ($\ell_1$ in Fig. \ref{Fig-flow-zoomJy}) along the tilted current sheet, although it is less accurate compared to the Hessian method. Angle of inclination calculated using the latter technique, for different shear flow cases, are listed in Table \ref{table3}. Likewise, $\ell_2$ is plotted perpendicular to $\ell_1$ denoting the direction across the CS. The length between intersecting points $a_1, a_2$ is estimated as the current sheet length. However, the $\B$ and $\vel$ field profiles in the downstream region appear to have different alignment with respect to $\ell_1$. Hence the exact location to measure the downstream field components is not clear. Moreover, in the upstream side, magnetic flux enters into the diffusion region along the line joining O-points ($\ell_3$). Hence the upstream $\B$ and $\vel$ components (denoted as $v^{up}_x, v^{up}_z, B^{up}_x, B^{up}_z$) are calculated at $b_1$ (or $b_2$) and plotted for different $v_0$ and $a_v$ values in Fig. \ref{Fig-flow-parameter-scaling}. Component $v^{up}_x$, measures the strength of secondary flow inside the island, increases with $v_0$. However, as the peak reconnection time increases with increasing shear flow strength (increasing $v_0$ and decreasing $a_v$ values, see Fig. \ref{Fig-recRate-ScanV0}), the effect of viscous dissipation also increases causing the $v^{up}_x$ value to decrease. In $v^{up}_z$ plot, its value for weak shear flow cases is of same order as that of no shear flow case (see Table \ref{table2} for comparison) and gradually increases with $v_0$ as the inclination of O-point joining line also increases generating a stronger $z$-component of secondary flow. At the same time, increasing effect of viscosity is also clearly visible as $a_v$ decreases. The behavior of $B^{up}_x$ values in different shear flow cases is explained in the previous paragraph. Likewise, in $B^{up}_z$ plot, the role of O-point joining line titling angle and viscosity values are similar to that of $v^{up}_z$ case. Half-length value of the current sheet also varies in a similar way as the peak $E_y$ value and $B^{up}_x$ do. It remains almost constant up to certain values of $v_0$: $v_0 = 1.0v_A, 0.7v_A, 0.5v_A, 0.3v_A$ for four $a_V$ cases. As the peak $E_y$ time increases due to flow driven reconnection, the size of islands at the beginning of coalescence phase become smaller generating a CS of smaller length. This island size-dependent behavior of CS length is similar to that observed in the previous subsection for no-shear flow case in Fig \ref{Fig-4}. Finally, the half-width of the current sheet is observed to increase with shear flow strength. . 
%%
%\begin{figure*}[!h]
%	\begin{subfigure}{0.45\textwidth}
%		\includegraphics[scale=0.23]{FIG_recRate-time_av-2aB}
%		\caption{}
%		\label{2a}
%	\end{subfigure}
%	\begin{subfigure}{0.45\textwidth}
%		\includegraphics[scale=0.23]{FIG_recRate-time_av-1aB}
%		\caption{}
%		\label{2b}
%	\end{subfigure}
%	\begin{subfigure}{0.45\textwidth}
%		\includegraphics[scale=0.23]{FIG_recRate-time_av-0p5aB}
%		\caption{}
%		\label{2c}
%	\end{subfigure}
%	\begin{subfigure}{0.45\textwidth}
%		\includegraphics[scale=0.23]{FIG_recRate-time_av-0p25aB}
%		\caption{}
%		\label{2d}
%	\end{subfigure}
%%	\begin{subfigure}{0.45\textwidth}
%%		\includegraphics[scale=0.23]{FIG_recRate-scanV0_scanAv}
%%		\caption{}
%%		\label{2e}
%%	\end{subfigure}
%%	\begin{subfigure}{0.45\textwidth}
%%		\includegraphics[scale=0.23]{FIG_t-maxRate-scanV0-scanAv}
%%		\caption{}
%%		\label{2f}
%%	\end{subfigure}
%\end{figure*}
%
%\section{Island coalescence without shear flows: benchmark}
%\begin{figure}[!h]
%	\includegraphics[height=0.5,width=0.5]{}
%\end{figure}
%
\section{Summary} \label{summary}
{Using a 2D incompressible viscoresistive MHD model, here we report on the quantitative scaling of different reconnection parameters such as peak $E_y$, peak $E_y$ time, total coalescence time, upstream/downstream magnetic fields and velocity field components, CS half-width and half-length in two different parametric regimes of IC problem. In the first part, we discuss the effect of island size controlled by the parameter $\epsilon$ which is intrinsic to Fadeev equilibrium, keeping the system size controlling parameter $a_B$ fixed. We initialize the velocity components to zero. Then we calculate the above mentioned parameters for different island sizes. In the second part, we initialize an in-plane shear flow parallel to the asymptotic magnetic field lines of the Fadeev equilibrium and calculate scaling of the above said parameters for a wide range of shear amplitude and shear lengths. In both the cases, an initial perturbation is added to the vector potential to start the coalescence instability. In the case of resistive IC (in absence of shear flow), peak $E_y$ versus $\hat{\eta}$ scaling follows SP type (see Fig. 3b from Ref. [25] of original Manuscript) and is valid of different $\epsilon$ values as well (see Fig. 10 from Ref. [17] of original Manuscript). \\
A detailed comparison of our results from the first part of this paper with that of SP-model are as follows.}
\begin{itemize}
	\item {In SP model, the mass conservation ratio is $v_{in}l/v_{out}\delta \simeq 1 \implies v_{in}/v_{out} = \delta/l$ and energy conservation ratio is $B_{in}^{2}/v_{out}^{2} \simeq 1 \implies v_{out}= B_{in} = v_A^{in}$. In this study, we find that the mass conservation ratio holds good whereas the energy conservation ratio $B_{in}^{2}/v_{out}^{2} \simeq 2 \implies v_{out} \simeq 0.7B_{in} ~\text{or}~ v_{out} \simeq 0.7 v_A^{in}$ (as seen in Table \ref{table2}). Further, the energy conservation ratio remains constant for all the $\epsilon$ values. This validate the correctness of our calculations.}
	\item With increasing island width, the magnitude of $J_y$ inside the islands and hence the strength of the Lorentz force between the islands increases. In larger islands, magnetic pressure pile-up at earlier times, is evident in both time varying reconnection rate and O-point tracing plots. We observe the magnitude of magnetic pressure to be decreasing, resulting in a decrease of total coalescence time for larger islands {i.e. the coalescence process is faster in larger islands}.
	\item The unnormalized peak $E_y$ increases monotonically with island width where as the normalized peak $E_y$ values (normalized to $v_A^{in} B_{in}$) are found to be weakly dependent on the $\epsilon$ values. 
	\item With increasing island width, the $B_{in}$ value increases due to increased flux content of islands, $v_{out}$ increases {and $v_{out} < v_A^{in}$ remains valid}. Likewise, $B_{out}$ and CS length $l$ values remain independent of island size. $v_{in}$ and CS width $\delta$ increases with $\epsilon$ value. 
	\item {We found the value of peak $E_y$ is not equal to the ratio of either $v_{in}/v_{out}$ or $\delta/l$. These surprising deviations in MIC problem from the Sweet-Parker model is also observed in Ref \cite{knoll2006POP}, although they have not explicitly reported therein. One possible explanation to these deviations is discussed as follows. In the IC problem, the finite sized islands supply the magnetic flux to the reconnection process and that decides the CS length. However, SP model assumes the magnetic flux supplied to the reconnection process by an infinite flux source. Although, a 1D calculation to this finite geometric effect is discussed in \cite{simakov2006}, further work required to clarify this.}
\end{itemize}
In the shear flow affected magnetic island case, the following are our observations.
\begin{itemize}
	\item We observed two different flow regions, distinguished by their flow patterns (see Fig. \ref{Fig-flow-zoomJy}): primary shear flow dominant region having shear flow parallel to asymptotic magnetic field and secondary shear flow region lying inside the islands. The secondary flow is induced due to the primary shear flow, having magnitude proportional to $v_0$. 
	\item {The total coalescence time is observed to have two different phases. First, there is a flow dominant phase where the shear flow tries to peel off the island making it difficult to distinguish the islands and a part of the magnetic flux get reconnected at multiple locations. As time progresses, the strength of the shear flow decreases due to viscosity, the islands regain their shape; this is marked as start of the second phase or coalescence phase. In case of stronger primary flows (with higher $v_0$ or smaller $a_v$ cases) flow dominant phase lasts for longer time and during which more amount of flux that get reconnects. Hence, the amount of flux remains inside the islands is less compared to that of initial time.} This is the reason why peak $E_y$ decreases and peak $E_y$ time increases with increasing shear flow strength (increasing $v_0$ and decreasing $a_v$). 
	\item Stronger shear flow makes the CS, diffusion region and O-point joining line to tilt with respect to the $x-z$ coordinate. For comparatively weaker shear flows, the inclination angle of CS and diffusion region is along the clockwise direction of positive $x$-axis. As the shear flow becomes stronger, the CS inclination angle increases along the anti-clockwise direction where as the diffusion region remains inclined to the direction opposite to that of CS. 
	\item We have calculated the CS thickness and magnetic field/velocity field components at its edge along the upstream direction. Importantly, we have found the CS thickness to increase and $B_x$ to decrease with increase in flow strength. An approximate CS length is also calculated which seems to be controlled by the island size during the coalescence phase.
\end{itemize}
%
%{Magnetic island coalescence or interacting magnetic flux tubes has been detected in many in-situ observation in the Earth magnetosphere \cite{zhao2016coalescence,zhao2019} and interplanetary coronal mass ejection \cite{feng2019observations} events as well as in many laboratory experiments such as merging-compression start-up spherical tokamak \cite{browning2014,browning2015}. These observations about the reconnection event, the associated CS and orientation of different field quantities in the vicinity of the CS are strikingly homologous to our study. Hence this present study will be helpful in understanding these observations. Furthermore, in most of the previous island coalescence numerical simulation studies, the initial equilibrium is considered as non-Force-free unlike the real 2D/3D flux tubes. Previously reported work with force-free Fadeev equilibrium \cite{murtas2021coalescence,BirnHesse2007} have suggested a generation of out-of-plane flow that is believed to be affected in the presence of in-plane and out-of-plane external shear flow.}\\

{Coalescing flux tubes or magnetic islands of different sizes or length scales have been observed in many experiments and satellite observations such as in the magnetosphere \cite{wang2016coalescence, zhao2016coalescence, zhao2019}, interplanetary coronal mass ejections \cite{feng2019observations, Song2012}, observed indirectly during solar flares, merging-compression start-up experiments in spherical tokamak \cite{browning2014,browning2015}, etc. In most of these reports, the width of the flux tubes are of the order of ion skin depth. However, here we use a more simplified incompressible MHD model, that excludes several important aspects to make a comparison with the realistic scenarios. The assumption of incompressiblity is more appropriate when there is a strong guide field or mean field present in the plasma as in a tokamak configuration. Magnetic islands in the presence of in-plane and (or) out-of-plane shear flow are also commonly observed in laboratory experiments, for example, tearing mode islands in presence of zonal flows or mean flow \cite{chandra2015} as well as in the earth magnetosphere \cite{ma2016flow}. Further, past studies with finite compressible effects have shown the formation of slow-mode shock during the coalescence/reconnection process \cite{ma2016flow,murtas2021coalescence}. Study of these shock structures in the presence of super-Alfv\'enic and super-sonic flows could lead to a more complex phenomenon. Moreover, in the shear flow case, we are unable to calculate the downstream components where the primary flow is strong enough to affect the downstream flows, due to flow dynamics.} A theoretical calculation similar to one given in Ref. \cite{cassak2011} could help in better understanding of our results. Furthermore, the physics of Hall term and finite-Larmor-radius (FLR) effects has been ignored in this work, although these can be important in the case of collisionless plasmas (for ex. plasmas found in the earth magnetosphere and fusion devices \cite{mahapatra2020gyro}). This is an ongoing study and further findings will be reported in a future communication.}

\section*{Acknowledgment}
The simulations are performed on the Antya cluster at the Institute for Plasma Research (IPR). AS is grateful to the Indian National Science Academy (INSA) for the position of an Honorary scientist. We thank Dr A. Chattopadhyay, IPR for his valuable inputs.
\appendix
\renewcommand\thefigure{\thesection.\arabic{figure}}

\section*{Data Availability}
The data that support the findings of this study are available from the corresponding author upon reasonable request.
%
%\bibliographystyle{apsrev4-2}
%\section*{References}
% \bibliography{Revised_Manuscript_RGanesh}
 
%\begin{thebibliography}{1}
%\section*{References}
\bibliographystyle{apsrev4-1}
\bibliography{Manuscript.bib}

\end{document}